\documentclass[3p,number,preprint]{elsarticle}

\usepackage[english]{babel}
\usepackage{graphicx}
\usepackage{amsmath, amsthm, amssymb, amsfonts}
\usepackage{url}

\begin{document}

\title{SNEG - Mathematica package for symbolic calculations with
second-quantization-operator expressions}

\author[rz]{Rok \v{Z}itko}
\ead{rok.zitko@ijs.si}

\address[rz]{J. Stefan Institute, Jamova 39, SI-1000 Ljubljana, Slovenia}

\date{\today}


\begin{abstract}
In many-particle problems involving interacting fermions or bosons,
the most natural language for expressing the Hamiltonian, the
observables, and the basis states is the language of the second-quantization
operators. It thus appears advantageous to write numerical computer
codes which allow the user to define the problem and the quantities of
interest directly in terms of operator strings, rather than in some
low-level programming language. Here I describe a Mathematica package
which provides a flexible framework for performing the required
translations between several different representations of operator expressions:
condensed notation using pure ASCII character strings, traditional
notation ("pretty printing"), internal Mathematica representation using
nested lists (used for automatic symbolic manipulations), and various
higher-level ("macro") expressions. The package consists of a collection of transformation
rules that define the algebra of operators and a comprehensive library
of utility functions. While the emphasis is given on the problems from
solid-state and atomic physics, the package can be easily adapted to
any given problem involving non-commuting operators. It can be used
for educational and demonstration purposes, but also for direct calculations
of problems of moderate size.
\end{abstract}

\maketitle

{\bf PROGRAM SUMMARY}

\begin{small}
\noindent
{\em Manuscript Title:} 
SNEG - Mathematica package for symbolic calculations with
second-quantization-operator expressions \\
{\em Author:}
Rok \v{Z}itko\\
{\em Program Title:}
SNEG \\
{\em Journal Reference:}
\\
{\em Catalogue identifier:}
\\
{\em Licensing provisions:}
GNU Public License\\
{\em Programming language:}
Mathematica\\
{\em Computer:}
any computer which runs Mathematica\\
{\em Operating system:}
any OS which runs Mathematica\\
{\em RAM:}
problem dependent\\
{\em Number of processors used:}
1\\
{\em Supplementary material:}
{\em Classification:}
2.9 Theoretical methods, 5 Computer algebra, 6.2 Languages\\
{\em External routines/libraries:}
\\
{\em Subprograms used:}
\\
{\em Nature of problem:} 
Manipulation of expressions involving second quantization operators
and other non-commuting objects. Calculation of commutators,
anticommutators, expectation values. Generation of matrix
representations of the Hamiltonians expressed in the second quantization
language.\\
{\em Solution method:}
Automatic reordering of operator strings in some well specified
canonical order; (anti)commutation rules are used where needed.
States may be represented in occupation-number representation.
Dirac bra-ket notation may be intermixed with non-commuting operator
expressions.\\
{\em Restrictions:}
For very long operator strings, the brute-force automatic reordering becomes
slow, but it can be turned off. In such cases, the expectation values
may still be evaluated using Wick's theorem.\\
{\em Unusual features:}
SNEG provides the natural notation of second-quantization operators
(dagger for creation operators, etc.) when used
interactively using the Mathematica notebook interface.\\
{\em Additional comments:}
\\
{\em Running time:}
problem dependent\\
\end{small}

\newcommand{\vc}[1]{\boldsymbol{#1}}
\newcommand{\ket}[1]{\left| #1 \right\rangle}
\newcommand{\bra}[1]{\left\langle #1 \right|}
\newcommand{\vev}[1]{\left\langle #1 \right\rangle}
\newcommand{\braket}[1]{\langle #1 \rangle}

\newcommand{\mma}[1]{{\tt #1}}

\bibliographystyle{unsrt} %

\section{Introduction}

Computational science has emerged as the third paradigm of science,
complementing experiments and theory. Computers are now used to 
realistically simulate physical systems which are not accessible to
experiments or would be simply too expensive to study directly. They
also allow numerical treatment of theoretical models which cannot be
solved by analytical means nor by simple approximations. In this
field, it is still common practice to quickly write ad-hoc computer
codes for performing calculations for specific problems. In these
rapidly developed computer programs the problem definition and the
quantities of interest are typically hard-coded using the same
low-level programming language which is also used to implement the
method of solution. In more technical terms, the problem-domain and
the solution-domain languages tend to coincide. As the scientific
interests change with time, such codes often undergo successive
modifications and adaptations, often leading to maintainability issues
or even bugs. In software engineering, the proposed solution to such
difficulties is to use a domain-specific language (DSL), i.e., a
specification language adapted to a particular problem domain. Using a
DSL, the problem can be expressed significantly more clearly than
allowed by low-level languages. In the field of many-particle physics,
such a language already exists: the language of strings of
second-quantization operators (particle creation and annihilation
operators) in terms of which it possible to express the problem (the
Hamiltonian), the quantities of interest (the observables), and the
domain of definition (the basis states defined by the creation
operators applied to some vacuum state). Using an appropriate notation
is equally important: the operators are usually expressed as
single-character symbols, possibly with further indexes, and a dagger
is used to distinguish creation from annihilation operators. The
computer algebra system Mathematica makes it possible to both easily
define the DSLs and to establish a suitable notation for these DSLs.
In this article I describe package SNEG, which implements a DSL for
second-quantization expressions and provides the corresponding natural
notation for its output and several syntactically different but
semantically equivalent ways for entering the input expressions. 
In addition to facilitating the representation of the input
to numerical codes, the package is powerful enough to perform some
calculations directly (e.g., evaluation of the expectation values
using Wick's theorem, calculation and simplification of operator
commutators, etc.).

This paper is structured as follows. Section~\ref{sec1} is devoted to
the specification of basic elements (operators), their concatenation
(non-commutative multiplication) and their automatic reordering
(according to the canonical commutation/anticommutation rules or some
other specification); it also introduces the Dirac bra-ket notation
which can be mixed with the second-quantization operator expressions,
and the occupation-number-representation vectors. Section~\ref{sec2}
presents some examples of higher-level routines for generating
second-quantization expression (particle number and spin operators,
etc.) and their manipulation (commutators and anticommutators, etc.).
Section~\ref{sec3} details the utility routines for generating basis
states which satisfy chosen symmetries (particle number conservation,
rotational invariance in the spin space) as well as the routines for
generating the matrix representations of operator expressions in given
basis space; these routines are crucial for the applications of SNEG
as an input preprocessor for lower-level numerical computer codes. The
focus of Section~\ref{sec4} are symbolic sums with dummy indexes and
their automated simplification using pattern matching. Finally,
Section~\ref{sec5} describes the successful use of SNEG in the numerical
renormalization group package ``NRG Ljubljana''.

The package SNEG comes with detailed documentation which integrates in
the Mathematica interactive help system. Each SNEG function is
carefully documented and examples are provided. For this reason, the
function calls are not described in this article; instead,
the focus here is on the basic concepts, design choices, conventions
followed, and some applications.

SNEG is released under the GNU Public License (GPL) and the most
recently updated version is available from
\url{http://nrgljubljana.ijs.si/sneg}. The package comes with a
standard battery of test cases which may be used as a regression test,
but also to verify that the possible user's custom extensions do not
interfere with the expected behavior of the library.

While there are other packages for symbolic calculations with
non-commuting objects for Mathematica and for other computer algebra
systems (supercalc \cite{supercalc}, ccr\_car\_algebra \cite{ccrcar},
NCAlgebra \cite{ncalgebra},
NCComAlgebra \cite{nccomalgebra}, grassmann.m \cite{grassmann},
grassmannOps.m \cite{grassmannOps}), none appears to have the scope 
of SNEG. Furthermore, the goal of SNEG is different from
more specialized symbolic manipulation packages such as TCE \cite{tce,
hirata2006} for performing many-body perturbation theory in quantum
chemistry or FormCalc \cite{formcalc} for calculations in theoretical
high-energy physics. Instead, SNEG is principally intended to provide
a general framework in which more sophisticated solutions can be
implemented or as a tool that provides a more natural interface to the
user.

\section{Foundations}
\label{sec1}

The cornerstone of SNEG is a definition of non-commutative
multiplication with automatic reordering of operators in some standard
form (usually the conventional normal ordering with creation operators
preceding the annihilation operators) which takes into account
selected (anti)commutation rules. Use of the standard form reordering
allows automatic simplifications of expressions. 

\subsection{Operator objects and numeric objects}

In SNEG, operators are internally represented as Mathematica expressions (lists)
with a chosen head (typically a single-letter symbol) and containing
the necessary indexes as list elements, for example
\begin{equation}
\mma{a[\,]},\quad
\mma{c[k,sigma]}.
\end{equation}
The symbols need to be explicitly declared before they are used.
The declaration routines define the default
(anti)commutation properties of the objects. They also establish the
natural on-screen notation (``pretty-printing'') when the package is
used interactively with the Mathematica notebook interface. Both the
(anti)commutation properties and the pretty-printing can be 
modified according to user's requirements. For operators declared
to be bosons or fermions, the first element of the list (i.e., the
first ``index'') has a special role: it distinguishes creation
operators (\mma{CR}=0) and annihilation operators (\mma{AN}=1):
\begin{equation}
\mma{c[CR,k]} \to c^\dag_{k}, \quad
\mma{c[AN,k]} \to c_{k}.
\end{equation}
In addition, for fermions, by default SNEG follows the convention that
the last index is interpreted as the particle spin
(\mma{DO}=$\downarrow$=0 and \mma{UP}=$\uparrow$=1); this convention
is used when generating operator expressions using higher-level
functions (see below) and when pretty-printing the expressions on 
computer display:
\begin{equation}
\mma{c[CR,k,UP]} \to c^\dag_{k\uparrow}, \quad
\mma{c[CR,k,DO]} \to c^\dag_{k\downarrow}.
\end{equation}

In addition to the internal Mathematica representation and the pretty
printing, SNEG supports a third way of expressing operators using a
condensed notation in pure ASCII strings. Such strings start with
the operator symbol, followed by a + sign in case of creation operators,
then the arguments follow in the parenthesis. For example, the three
following expressions are equivalent:
\begin{equation}
\mma{c}\!+\!\mma{(k)} \quad \leftrightarrow \quad
\mma{c[CR,k]} \quad \leftrightarrow \quad
c^\dag_{k}.
\end{equation}
The translations between ASCII strings and Mathematica expressions
need to be performed explicitly using suitable functions, while pretty
printing is (by default) performed automatically to render an
expression in internal representation using the conventional notation.
In principle, it would be possible to use the conventional
pretty-printed notation for input, but entering such expressions by
hand or with the assistance of Mathematica palettes turns out to be
cumbersome.

Fermionic operators with different symbols are assumed to anticommute,
bosonic operators with different symbols are assumed to commute, and
bosonic and fermionic are assumed to commute. If necessary, this
default behavior can be overridden.

SNEG allows to explicitly declare certain symbols to be numeric
quantities of specific kinds (integers, real, complex, or Grassman
numbers); this information is used to correctly
factor out numeric objects from the operator strings. 
Grassman variables are correctly anticommuted and $z^2=0$.

\subsection{Non-commutative multiplication}

In SNEG, the non-commutative multiplication is internally 
denoted by \mma{nc}. In interactive sessions \mma{nc} multiplications are pretty printed with 
centered dots between the terms, for example:
\begin{equation}
\mma{nc[c[CR,k,UP], c[AN,k,UP]]} \to c^\dag_{k\uparrow} \cdot
c_{k\uparrow}.
\end{equation}
The dot is displayed in order to permit easy detection of possible
errors arising from an inadvertent replacement of non-commutative
with the usual commutative multiplication. 
Function \mma{nc} is linear in all its
arguments
and associative.
Furthermore, \mma{nc[\ ]}=1 and \mma{nc[c]}=\mma{c}; these two rules
mimic the behavior of the standard Mathematica product function
\mma{Times} and imply the property of idempotency of multiplication.
Function \mma{nc}, unlike \mma{Times}, does not have the
pattern-matching attributes \mma{Flat} and \mma{OneIdentity}. Instead,
the associativity property is explicitly implemented. This design
choice was motivated by reasons of efficiency in pattern matching, and
benchmarking has shown that the explicit rules perform better than the
version using \mma{Flat} and \mma{OneIdentity} by up to 50\%.

It is possible to convert Mathematica representation of operator
expressions to (and from) ASCII strings. In ASCII strings, the
non-commutative multiplication is implied. For example,
$\mma{a}\!+\!\mma{(k)a(l)}$ is converted to $\mma{nc[a[CR,k],a[AN,l]]}$.

\subsection{Expression reordering}

Computer algebra systems simplify expressions by ordering them in some
canonical manner; in this way, the equivalent parts can be combined (or
canceled out, if the prefactors sum to zero). This is how some
simplifications are automatically effected in Mathematica. For this
reason, SNEG also attempts to reorder multiplicands in the \mma{nc}
operator strings according to some canonical order, using the
associated canonical anticommutation and commutation rules for the
operators.

By default, fermionic operators are sorted canonically (in the sense
that creation operators are permuted to the left and annihilation
operators to the right)
and then by
the remaining indexes, including spin as the last index.  It has to be
remarked, however, that the canonical order depends on the definition
of the vacuum state. SNEG supports either an ``empty band'' vacuum
with no particles present, or a ``Fermi sea'' vacuum with levels
filled up to the Fermi level.
In the default "empty band" ordering the value of
the first index (\mma{CR} or \mma{AN}) fully determines whether an
operator is a creation or an annihilation operator.
In the case of
``Fermi sea'' ordering, the second index of the operator is tested by
default. This index is assumed to be a ``momentum'' or ``energy''
index, with the Fermi level fixed at zero. 
If necessary, it
is also possible to turn off the automatic reordering for a fermionic
operator.
This is useful for
very long expressions which can be simplified more efficiently by
explicitly using Wick's theorem rather than by
automatic operator reordering.

Internally, the operator ordering is tested with \mma{snegOrderedQ},
while the necessary transformations on the operator strings (when
out-of-order parts are detected) are implemented as transformation
rules for the function \mma{nc}. 

When operators of different types appear in the same product, they are
disentangled. For example, bosonic operators are by default always
commuted to the left of fermionic operators, Majorana fermionic
operators are anti-commuted to the left of the Dirac fermionic
operators, etc.

SNEG will attempt to simplify expressions involving exponential
functions of operators using the Baker-Hausdorff relations
and the Menda\v{s}-Milutinovi\'c relations \cite{mendas1989}.

\subsection{Occupation-number representation}

For fermionic operators, SNEG allows working with the
occupation-number representation (ONR) of the states in a given Fock
space. Second-quantized expressions can be applied to these states,
one can compute the matrix elements of operators between pairs of
states, etc.
The states in the ONR are expressed in the form of Mathematica lists
(``vectors'') with head \mma{vc}, which contain the occupancies of all
orbitals represented by zeros and ones.
In interactive
Mathematica notebooks, the ONR vectors are shown in the Dirac-ket-like
format with boxes which are either empty or filled, according to the
occupancy of various orbitals:
\begin{equation}
\mma{vc[0,1,0,1]} \to \ket{\square \blacksquare \square \blacksquare}.
\end{equation}
If a vector is conjugated using \mma{conj}, it behaves as the
corresponding bra. If a bra and a ket are multiplied by \mma{nc}, the
corresponding scalar product is computed.
It is possible to convert an ONR vector to the string of creation
operators which, applied to the vacuum state, would give back the same
vector.

\subsection{Dirac's bra-ket notation}

SNEG provides support for calculations with the Dirac bra-ket
notation, which can be intermixed with the second-quantization
expressions. This is convenient, for example, for mixed
electron-phonon systems, where the fermions can be described using the
second-quantization language, but the oscillator using some other
convenient representation, such as coherent states. In mixed
expressions, the bras and kets are by default always commuted to the
right of the fermionic operators. In interactive Mathematica sessions,
the bras and kets are displayed enclosed by appropriate angled
brackets and bars.

A ket can be expressed using function \mma{ket}, which can take one or
several arguments (quantum numbers):
\begin{equation}
\mma{ket[m,n]} \to \ket{m,n}.
\end{equation}
Quantum numbers may also remain unspecified; this is signaled by the
value \mma{Null}; in interactive sessions it is displayed as a small
centered circle:
\begin{equation}
\mma{ket[m,Null,n]} \to \ket{m,\circ,n}.
\end{equation}
This functionality can be used to multiply kets from orthogonal
Hilbert spaces.
All the preceding rules also apply to bras, defined using \mma{bra}.
With the Hermitian conjugation function \mma{conj}, a bra can be
transformed into ket and vice versa. 

When a bra and a ket are multiplied by \mma{nc}, a scalar product is
computed by comparing the quantum numbers in equivalent positions
using the Kronecker delta:
\begin{equation}
\mma{nc[bra[m,n],ket[i,j]]} \to \delta_{m,i} \delta_{n,j},
\end{equation}

It is possible to mix occupation-number-representation vectors and
Dirac bra-ket vectors. The two subspaces are assumed to be unrelated
(i.e., tensor product space). 

\section{Generation of expressions and operations on expressions}
\label{sec2}

SNEG includes higher-level functions for generating various physically
relevant operators which can be expressed in terms of the
second-quantization operators and for performing various operations
upon the expressions. In many of the applications of SNEG, the package
can be used at this higher level and the user does not need to be
concerned with the inner working of the library.

\begin{itemize}

\item
The number (occupancy) operator $n=c^\dag c$ can be generated with the
function \mma{number} which comes in different flavors depending on
the function argument(s). It can automatically handle particles with
spin and it is possible to generate the number operator for more complex
objects such as linear combinations of orbitals.

\item
The inter-site hopping operator may be generated using \mma{hop},
for example for a particle with spin:
\begin{equation}
\mma{hop[c[1], c[2]]} \to \sum_\sigma c^\dag_{1,\sigma} c_{2,\sigma}
+ c^\dag_{2,\sigma} c_{1,\sigma}.
\end{equation}

\item

The electron-electron repulsion operator $n_\downarrow n_\uparrow$ may
be generated using the SNEG function \mma{hubbard}:
\begin{equation}
\mma{hubbard[c]} \to -c^\dag_\downarrow c^\dag_\uparrow c_\downarrow
c_\uparrow.
\end{equation}

\item

Spin operator for orbitals described by fermionic operators can be
generated by SNEG functions \mma{snegx}, \mma{snegy}, and \mma{snegz},
which are defined for all values of particle spin (parameter
\mma{spinof}). For example,
\begin{equation}
\mma{spinx[c]} \to 
\frac{1}{2} \left( c^\dag_\downarrow c_\uparrow + c^\dag_\uparrow
c_\downarrow \right)
\end{equation}
for a spin-$1/2$ operators.
The exchange coupling (i.e., the scalar product of two spin operators,
$\vc{S}_1 \cdot \vc{S}_2$) can be generated using
\mma{spinspin}.

\item

An important application area of SNEG is the computation of the vacuum
expectation values (VEV) of second-quantization-operator strings using
\mma{vev}.
To speed up the evaluations, a number of simplification rules are
defined in SNEG. 
Expressions can be ``normal ordered'' by subtracting their vacuum
expectation values.

\item

Hermitian conjugates of operator strings can be computed using the
function \mma{conj}. The numerical constants and parameters are
handled correctly depending on their nature (real or complex commuting
numbers, or Grassman anticommuting numbers). For complex fermionic and
bosonic operator objects, the first index is modified (creation to
annihilation, and vice versa), while real fermionic operator objects
are left unchanged.

\item

Commutators and anticommutators can be computed trivially by forming
the sums or differences of the products, or with the help of the
provided auxiliary functions \mma{commutator} and
\mma{anticommutator}. 

\item

Projection operators can be generated, for example
$\mma{projector0[c]} \to (1-n_\uparrow)(1-n_\downarrow)$.

\end{itemize}
This list is not exhaustive and further functionality is described
in the bundled SNEG documentation.

Often it is necessary to proceed in the opposite direction: given a
long complex operator-string expression, one has to rewrite it in
terms of higher-level functions, such as number, hopping, repulsion,
or spin operators. This might be used, for example, after performing a
change-of-basis transformation on the creation and annihilation
operators, if one requires a physical interpretation of the resulting
long expression. There is, clearly, no unique mapping from an expression to
the corresponding generation functions. Therefore, there are several
specialized SNEG routines which apply heuristic rules in an attempt to
rewrite the expression.
The whole
set of the rules can also be applied by \mma{SnegSimplify} and
\mma{SnegFullSimplify}, although experience shows that such
brute-force approach is not very efficient and that a guided
consecutive application of suitably chosen specialized routines gives
better results.

\section{Generation of sets of basis states}
\label{sec3}

One of the principal application areas of SNEG is the transformation
of the operators expressed in the second-quantized notation into the
corresponding matrix representation in a given Hilbert space. To
simplify numerics, it is often important to take into account various
symmetries of the problem, i.e., to determine the Hamiltonian matrices
in the different invariant Hilbert subspaces. SNEG provides a number
of functions for generating the basis-state sets with chosen
well-defined quantum numbers (total charge, total spin, etc.).

\begin{figure}[htbp]
\centering
\includegraphics[width=10cm]{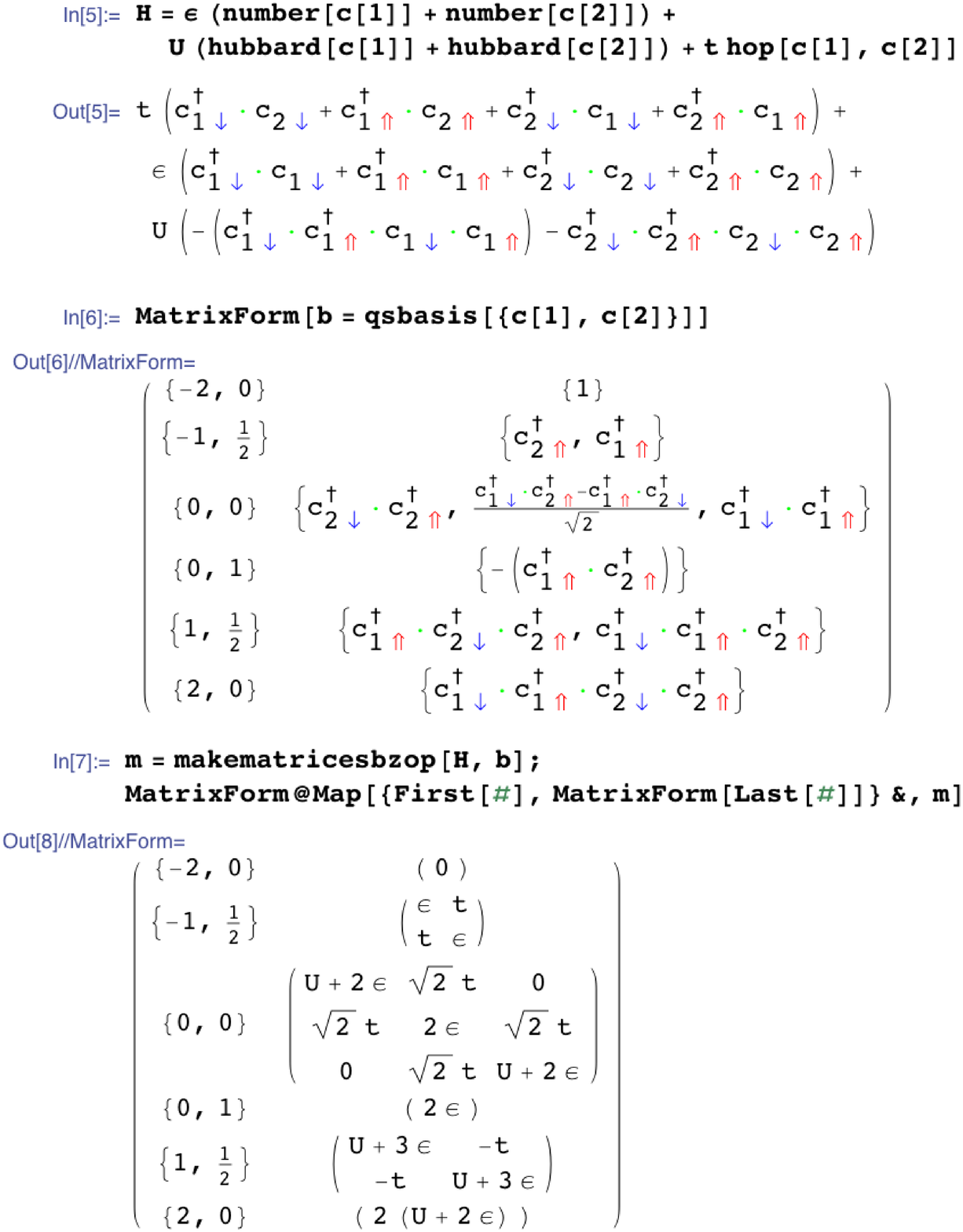}
\caption{A screen-shot from an interactive SNEG session using
the Mathematica notebook interface: definition of the Hamiltonian
$H$ for a two-site Hubbard model, generation of a basis with
well-defined total occupancy (relative to the half-filling) and
total spin, and conversion of the Hamiltonian to its matrix
representations in each invariant subspace of the total Fock
space. The index pairs in the first column of the table are
the quantum numbers, for example $\{1,1/2\}$ corresponds to
$Q=1$ (one particle above half-filling) and $S=1/2$ (i.e.
a spin-doublet state).}
\label{fig}
\end{figure}

The states can be represented in SNEG either
as strings of second-quantization operators (implicitly applied to a
vacuum state in which no particles are present) or as
occupation-number-representation vectors. 
The basis-state set that spans the full Fock space is represented as a
list of pairs -- the first member of each pair is a list of quantum
numbers which fully characterizes the invariant subspace, while the
second member of the pair is a list of all the basis states in the
given subspace. 
Once the desired basis sets have been generated, the operators in the
second-quantization language can be transformed into the corresponding
matrix representations. An example of such a calculation for the
two-site Hubbard model is shown in Fig.~\ref{fig} in the form of an
interactive Mathematica session. One can see how easy it is to extend
such a model definition to larger cluster sizes, to add various
interaction terms to the Hamiltonian, or to define basis sets which
satisfy different symmetries. For example, in the presence of the
magnetic field (added using the macro function {\mma{spinz}}), one
should replace the full spin quantum number $S$ by a component of spin
along the $z$-axis, $S_z$; such a change is effectuated by replacing
the call to {\mma{qsbasis}} by a call to {\mma{qszbasis}}, no other
changes are required. This example makes it clear how easy it is to
add various perturbation terms to the Hamiltonian; very often no
programming is required.

\section{Symbolic sums}
\label{sec4}

SNEG allows calculations with symbolic sums over dummy indexes, which
remain in their unevaluated forms. They are defined using the
function \mma{sum} taking two arguments: the first one is the
expression that is being summed over, while the second one is the list
of all summation indexes. 
The list of indexes is automatically sorted; this allows some
automatic simplifications. Numeric quantities which do not depend on
any of the summation indexes are factored out.
Products of sums can be calculated using \mma{nc}. SNEG automatically
handles summation index collisions and renames the duplicated indexes.
It is thus perfectly safe to use the same dummy index in different
expressions.
When commutators of sums are computed, the name replacement is always
performed on the same sum in order to maximize the opportunities for
automatic cancellation of equal terms.

Expressions with symbolic sums can be automatically
simplified using \mma{sumSimplify} which uses Mathematica function
\mma{Simplify} with additional transformation functions for
expanding, simplifying, and collecting the terms.

\section{Applications}
\label{sec5}

SNEG has found many applications in the field of
theoretical condensed-matter physics. It has been applied to perform
exact diagonalizations on Hubbard clusters, small Heisenberg chains,
and similar lattice models, calculations of commutators of complex
operator expressions (to establish the presence of various symmetries,
in the equation of motion method, etc.), and perturbation theory to
higher orders. It is best suited for problems where the complexity is
too high for paper-and-pencil calculations, yet still sufficiently low
for a brute-force computer algebra approach (which SNEG essentially
is). The package makes otherwise tedious calculations a routine
operation. Most importantly, it prevents inauspicious sign errors
which commonly arrise when fermionic operators are (anti)commuted. For
this reason, the package is also suitable for educational purposes,
i.e., as a way of verifying the correctness of elementary
calculations with operator quantities.

It should be remarked that the automatic-reordering approach does not
scale well to very long strings of operators, since the computation
time for automatic simplfications increases as a power law of the
string length. This is due to the fact that SNEG internally makes use
of Mathematica as a pattern matching and replacement engine. The
pattern application time increases with the string length, but also
the application of (anti)commutation rules leads to the rapid growth
of the total number of terms in the intermediate expressions before
the terms can be eventually cancelled out. For such problems, the
automatic reordering can be turned off; the expression can still be
reordered explicitly using a suitable direct calculation. For very
long operator strings, it would be more appropriate to perform the
expression reordering and simplification using a direct algorithm
coded, perhaps, in some lower-level programming language.

The main major application of SNEG in its role of an ``interface''
between the user and the low-level numerical codes is the
package ``NRG Ljubljana'' for performing the numerical renormalization
group (NRG) calculations for quantum impurity models
\cite{wilson1975, krishna1980a, bulla2008}. Using SNEG as the
underlying library, both the model (Hamiltonian) and the observables
(operators) may be defined in terms of high-level expressions. This
enables a clear separation between the problem domain (coded in
Mathematica and SNEG) and the solution domain (coded in C++). This is
advantageous not only for reasons of performance, but especially for
maintainability of the code. During the lifetime of the project, no
major rewrites or design changes were necessary in either part of the
code and the development could proceed incrementally without breaking
the existing features. Furthermore, adapting the package to different
problems and symmetries, or to calculate new quantities, is rather
trivial.

\section{Conclusion}

The natural language of many-particle physics is the quantum field
theory, more particularly the formalism of the second quantization. I
have argued that the computational many-body physics should strive
towards creating computer codes which allow defining problems --
whenever possible -- in their natural problem-domain language. It is
hoped that the approach (and the specific implementation, SNEG) will
improve the productivity of users and the quality of scientific
software, in particular reliability, reusability, maintainability, and
correctness.

\bibliography{paper}

\end{document}